# Journey from Data Mining to Web Mining to Big Data

**Richa Gupta**

*Department of Computer Science*
*University of Delhi*

**ABSTRACT :** *This paper describes the journey of big data starting from data mining to web mining to big data. It discusses each of this method in brief and also provides their applications. It states the importance of mining big data today using fast and novel approaches.*

**Keywords-** *data mining, web mining, big data*

## 1    INTRODUCTION

Data is the collection of values and variables related in certain sense and differing in some other sense. The size of data has always been increasing. Storing this data without using it in any sense is simply waste of storage space and storing time. Data should be processed to extract some useful knowledge from it.

## 2    DATA MINING

Data Mining is analysing the data from different perspectives and summarizing it into useful information that can be used for business solutions and predicting the future trends. Mining the information helps organizations make proactive, knowledge-driven decisions and answer questions that were previously time consuming to resolve.

Data mining (DM), also called Knowledge-Discovery in Databases (KDD) or Knowledge-Discovery and Data Mining, is the process of automatically searching large volumes of data for patterns such as association rules. It is a fairly recent topic in computer science but applies many older computational techniques from statistics, information retrieval, machine learning and pattern recognition.

Data mining is important as the particular user will be looking for pattern and not for complete data in the database, it is better to read wanted data than unwanted data. Data mining extract only required patterns from the database in a short time span

Based on the type of patterns to be mined, data mining tasks can be classified into summarization, classification, clustering, association and trends analysis [1].

**Summarization** is the abstraction or generalization of data. Data is summarized and abstracted to give a smaller set which provides overview of data and some useful information [1].

**Classification** is the method of classifying objects into certain groups based on their attributes. Certain classes are made by analyzing the relationship between the attributes and classes of objects in the training set [1].

**Association** is the discovery of togetherness or connection of objects. The association is based on certain rules, known as association rules. These rules reveal the associative relationship among objects, that is, they find the correlation in a set of objects [1].

**Clustering** is the identification of clusters or groups for a set of objects whose classes are unknown. Clustering should be done such that the similarities between objects of same clusters are maximized and similarities between different clusters are minimized.

**Trend analysis** is discovering of interesting patterns in the dimension of time. It is the matching of objects' changing trends such as increasing streaks [1].

Data Mining tools can be classified into three categories: traditional data mining tools, dashboards and text-mining tools.

- Traditional Data Mining Tools helps companies establish data patterns and trends using complex algorithms and techniques usually situated on a single computer. The majority are available in Windows and Unix versions. They normally handle data using offline tools [2].

- Dashboards reflect data changes enabling the user to see how the business is performing.
- Text mining tools mines data from different kinds of texts, example, Microsoft word, acrobat PDF, text files. These tools scan the content and convert the selected data into format that is compatible with tool's database [2].

## 2.1 APPLICATIONS OF DATA MINING

- Artificial neural networks.
- Business applications. In this it is used for database marketing, retail data analysis, stock selection, credit approval etc.
- Science applications. It is used in astronomy, molecular biology, medicine, geology etc.
- Health care management
- Tax fraud detection

## 3 WEB MINING

As the usage of web started to increase, so does the demand of data mining. Web mining is the application of data mining techniques to discover usage patterns from large Web repositories. It reveals interesting and unknown knowledge about both users and websites which can be used for analysis. It is used to understand customer behaviour, evaluate the effectiveness of a particular website and help quantify the success of a marketing campaign [3, 4]. Web mining can be classified into three types based on the type of data:

- **Web content mining** – it is the process of extracting useful information and knowledge from the web contents/data/documents. Content may consist of text, images, audio, video or structured records such as lists and tables. Web content mining is differentiated from two different points of view: Information Retrieval View and Database View [5].
- **Web structure mining** – it is the process of using graph theory to analyse the node and connection structure of a website. It tries to discover the underlying link structures of the web. It can be used to generate information on the similarity or the difference between different websites [4].
- **Web usage mining** – it attempts to discover useful knowledge from the data obtained from web user sessions. It tries to find usage patterns from the web data to understand and better serve the needs of Web-based applications. Some applications of web usage mining are adaptive websites, web personalization and recommendation, business intelligence.

## 3.1 APPLICATIONS OF WEB MINING

- It has its great use in e-commerce and e-services
- In e-learning
- Self-organizing websites
- Digital libraries
- E-government
- Security and crime investigation

## 4 BIG DATA

There has been a lot of growth in the amount of data generated by web these days. The data has been so large that it becomes difficult to analyse it with the help of our traditional mining methods. Big data term has been coined for data that exceeds the processing capability [6]. It has three main key characteristics

- Volume – the size of data is now larger than terabytes and petabytes. This large scale makes it difficult to analyse using conventional methods.
- Velocity – big data should be used to mine large amount of data within a pre-defined period of time. The traditional methods of mining may take huge time to mine such a volume of data.
- Variety – big data comes from various sources. It is designed to handle structured, semi-structured as well as unstructured data. Whereas the traditional methods were designed to handle structured data and that too not of such large volume.

Big data is a general term for massive amount of digital data being collected from various sources, that are too large and raw in form. Big data deals with new challenges like complexity, security, risks to privacy. Big data is redefining the data management from extraction, transformation and processing to cleaning and reducing [7].

## 4.1 APPLICATIONS OF BIG DATA

- In social networking sites to find for usage patterns
- In google search
- Astronomy
- Sensor networks
- Government data
- Web logs
- Mobile phones
- Natural disaster and resource management
- Scientific research

## 5 CONCLUSION

In this paper, we have reviewed the journey on how on big data evolved. It defines the traditional mining methods as data mining, then with the advancement of web, came the concept of web mining. And later on, the size and variety of data pushed us to think ahead and develop new and faster methods of mining data which uses the parallel computing capability of processors. This term is known as Big data. We have also provided with the applications of different methods of mining.